\documentstyle{article}
\begin{document}
\LARGE
\begin{center}
\bf  Can Black Holes be Created at the Birth of the Universe ?

\vspace*{0.6in}
\normalsize \large \rm 

Zhong Chao Wu

Dept. of Physics

Beijing Normal University

Beijing 100875, China

(Gravity Essay)

\vspace*{0.4in}
\large
\bf
Abstract
\end{center}
\vspace*{.1in}
\rm
\normalsize
\vspace*{0.1in}
We study the quantum creation of black hole pairs in the 
(anti-)de Sitter space background. These black hole pairs in the
Kerr-Newman family are created from constrained instantons. At
the
$WKB$ level, for the chargeless and nonrotating case, the
relative creation probability  is the exponential of (the
negative of) the entropy of the universe. Also for the remaining
cases of the family, the creation probability is the exponential
of (the negative of) one quarter of the sum of the inner and
outer black hole horizon areas. In the absence of a general 
no-boundary proposal for open universes, we treat the creations
of
the closed and the open universes in the same way.

PACS number(s): 98.80.Hw, 98.80.Bp, 04.60.Kz, 04.70.Dy

Keywords: quantum cosmology, constrained gravitational instanton,
black hole creation

\vspace*{0.3in}

e-mail: wu@axp3g9.icra.it

\pagebreak

\vspace*{0.3in}

There are three ways of forming black hole in Nature. The
first way is through the gravitational collapse of a massive body
in
astrophysics. In this scenario, the spacetime 
and matter content are treated classically. In general, the
effect of the cosmological background is ignored. The second way
originates from the quantum fluctuation of the
matter content in the  very early universe. Here the spacetime is
again treated classically. The black hole formation is a result
of  the competing effects of the expansion of the universe and
the gravitational attraction of the matter fluctuation.
The third way is through the quantum creation of black holes in
quantum cosmology, to which this paper is addressed. Here, both
the spacetime and the matter content are quantized. This is the
most dramatic type of black hole formation. Indeed, the black
holes are essentially created from nothing at the same moment as
the birth of the universe. Therefore, only black holes created
this way are genuinely primordial. 

It is believed that the Planckian era of the universe
underwent an inflationary stage which was approximated by the de
Sitter metric. In the Planckian stage, the potential of the
scalar field behaves as an effective cosmological constant
$\Lambda$. On the other hand, extended theories of
supergravity in which the $O(N)$ group is gauged have the anti-de
Sitter space as their ground or most symmetric state.  Therefore,
it is of great interest to study quantum creations of black holes
in these backgrounds.

In the No-Boundary Universe, the wave function of a closed
universe is  defined as
a path integral over all compact 4-metrics with matter fields
[1]. The dominant contribution to
the path integral is from the stationary action solution. At
the $WKB$ level, the wave function
can be approximated as $\Psi \approx e^{- I}$,
where $I= I_r + iI_i$  is the complex action of the solution.

The imaginary part $I_i$ and real part $I_r$ of the
action represent the Lorentzian
and Euclidean evolutions in real time and imaginary time,
respectively. The probability of a Lorentzian orbit remains
constant during its evolution. One can identify the probability,
not only as the probability of the universe created, but also as
the probabilities for other Lorentzian universes obtained through
an analytic continuation from it [2].

An instanton is defined as a stationary action orbit and
satisfies the Einstein equation everywhere. It was thought that,
at the $WKB$ level, an instanton was the seed for the creation of
the universe. Very recently, it was realized that this only
applied to the case of creation with a stationary 
probability.  Therefore, in order not to exclude many
interesting phenomena and more realistic models from
the study, one has to appeal to the concept of constrained 
instantons [3][4][5]. Constrained instantons are the orbits
with an action that is stationary under some restriction. The
restriction can be imposed on a spacelike 3-surface of the
created Lorentzian universe. The restriction is that the 3-metric
and matter content are given at the 3-surface. The relative
creation probability from the instanton  is the exponential of
the negative of the real part of the instanton action.

One can begin with  a complex solution to the Einstein equation
and other field equations in the complex domain of spacetime
coordinates. If an instanton exists at all, then it should be
a compact singularity-free section of the solution. If there are
singularities in the compact section, then, in general, the
action of the section is not stationary. The action may
only be stationary with respect to the variations under some
restrictions mentioned above. We call  this section  a
constrained gravitational instanton. To find the constrained
instanton, one has to closely investigate the singularities. The
stationary action condition is crucial to the validation of the
$WKB$ approximation, which we use to investigate 
the problem of quantum creation of a black hole pair.

In contrast to the case for a closed universe, a general
no-boundary proposal for the quantum state of  an open universe
has not been presented. However, one can use analytic
continuation from a complex constrained instanton to obtain the
$WKB$ approximation to the wave function for open universes with
some kind of symmetry. At this level, both the open and closed
creations of universes can be dealt with in the same way. For
examples, The $S^4$ space model with $O(5)$ symmetry [6] and the
$FLRW$ space model with $O(4)$ symmetry [2] have been
investigated this way.

The constrained gravitational instantons for the pair creation of
black holes in the
(anti-)de Sitter space background can be obtained from the
complex solutions of the Kerr-Newman-(anti-)de Sitter family [7]
\begin{equation}
ds^2 = \rho^2(\Delta^{-1}_r dr^2 + \Delta^{-1}_\theta d\theta^2)
+ \rho^{-2}
 \Xi^{-2}
\Delta_{\theta} \sin^2 \theta (adt - (r^2 + a^2) d\phi)^2 -
\rho^{-2} \Xi^{-2}\Delta_r  (dt - a \sin^2 \theta d \phi)^2,
\end{equation}
where
\begin{equation}
\rho^2 = r^2 + a^2 \cos^2 \theta,
\end{equation}
\begin{equation}
\Delta_r = (r^2 + a^2)(1 - \Lambda r^2 3^{-1}) - 2mr + Q^2,
\end{equation}
\begin{equation}
\Delta_{\theta} = 1 + \Lambda a^2 3^{-1} \cos^2 \theta,
\end{equation}
\begin{equation}
\Xi = 1 + \Lambda a^2 3^{-1}
\end{equation}
and $m, ma$ and $Q$ are constants,  representing
mass, angular momentum, electric or magnetic charges. We shall
not consider the dyonic case in the following. We shall
respectively call the cases with de Sitter and anti-de
Sitter backgrounds as closed and open models.

We use $r_0, r_1, r_2$ and $r_3$ to denote the four roots of
$\Delta_r$. For the closed model with positive $\Lambda$, we
assume
all roots $r_0, r_1, r_2$ and $r_3$ are real and in ascending
order. These roots are the negative,
inner black hole, outer black hole and cosmological horizons,
respectively. For the
open model with negative $\Lambda$, at least two roots, say $r_0,
r_1$, 
are complex conjugates, and we assume $r_2$ and $r_3$ are real.
If this is the case, then
$r_2$ and $r_3$ must be positive and can be identified as the
inner 
and outer black hole horizons, respectively.

For the closed model [3], the constrained instanton is
constructed from the  metric (3) 
by setting $\tau = it$. One makes two cuts at
$\tau = \pm \Delta \tau /2$ between the two  horizons
$r_1, r_2$ and glues them. The resultant manifold may have
conical singularities at the two horizons.  It has the $f_1$-fold
cover around the horizon $r_1$ and the  $f_2$-fold cover around
the horizon $r_2$.

The Lorentzian metric for the created black hole pair is obtained
through analytic continuation of the time coordinate from an
imaginary value to a real value at the equator. The equator is
two joint sections $\tau = consts.$ passing these horizons. It
divides the instanton into two halves. We can impose the
restriction that  the 3-geometry characterized by the parameters
$m, a$ and $Q$ is given at the equator for the Kerr-Newman-de
Sitter family. The parameter $\Delta \tau$ is the only
degree of freedom left for the pasted manifold, since the field
equation holds everywhere with the possible exception of these
horizons. Thus, in order to check whether we get a stationary
action solution for the given horizons, one only needs to see
whether the above action is stationary with respect to this
parameter. The equator where the quantum transition will occur
has the topology $S^2 \times S^1$.

The action due to the horizons is [3]
\begin{equation}
I_{i, horizon} = - \frac{\pi (r^2_i + a^2)(1 -
f_i)}{\Xi}.\;\; (i = 1,2)
\end{equation}
 
The action due to the volume is
\begin{equation}
I_v = - \frac{\Delta \tau \Lambda}{6\Xi^2} (r^3_2 - r^3_1 +
a^2(r_2 - r_1)) \pm \frac{\Delta \tau Q^2}{2\Xi^2} \left (
\frac{r_1}{r^2_1+ a^2} -
\frac{r_2}{r^2_2 + a^2} \right ),
\end{equation}
where $+(-)$ is for the magnetic (electric) case.

If one naively takes the exponential of the negative of half the
total action, then the exponential is not identified as the wave
function at the  creation moment of the black hole pair. The
physical reason is that what one can observe is only the angular
differentiation, or the relative rotation of the two horizons.
This situation is similar to the case of a Kerr black hole pair
in the asymptotically flat background. There one can only measure
the rotation of the black hole horizon from spatial infinity. To
find the wave function for the given mass and angular momentum
one has to make the Fourier transformation [3][8]
\begin{equation}
\Psi(a, h_{ij}) = \frac{1}{2 \pi}\int^{\infty}_{-\infty}
d\delta e^{i\delta
J \Xi^{-2}} \Psi(\delta, h_{ij}),
\end{equation}
where $\delta$ is the relative rotation angle for the half time
period $\Delta \tau /2$, which is canonically conjugate to the
angular momentum $J = ma$; and the factor $\Xi^{-2}$ is due to
the time rescaling. The angle difference $\delta$ can be
evaluated
\begin{equation}
\delta = \int_0^{\Delta \tau/2} d\tau (\Omega_1 - \Omega_2),
\end{equation}
where the angular velocities at the horizons are
$\Omega_i = a(r^2_i + a^2)^{-1}$.

In the magnetic case the vector potential determines the magnetic
charge, which is the integral  over the $S^2$ factor.
However, in the electric case, one can only fix the integral
\begin{equation}
\omega = \int A,
\end{equation}
where the integral is around the $S^1$ direction, and $A$ is the
vector potential of the electric field [3].
So, what one obtains in this way is $\Psi(\omega, a, h_{ij})$.
However, one can get the wave function $\Psi (Q,a, h_{ij})$ for
a given electric charge through the Fourier transformation
[3][8][9][10]
\begin{equation}
\Psi (Q,a, h_{ij}) = \frac{1}{2\pi} \int^{\infty}_{-\infty} d
\omega e^{i\omega Q} \Psi
(\omega,a, h_{ij}).
\end{equation}

The Fourier transformations (8) and (11) for the angular
momentum and
the electric charge are equivalent to adding  extra terms into
the action for the constrained instanton, and then the total
action becomes [3]
\begin{equation}
I = - \pi(r^2_1 + a^2)\Xi^{-1} - \pi(r^2_2 + a^2)\Xi^{-1}.
\end{equation}

It is crucial to note that the action is independent of the time
identification period $\Delta \tau $ and therefore, the manifold
obtained is qualified as a constrained instanton.
The relative probability of the Kerr-Newman black hole
pair creation from the constrained instanton is
\begin{equation}
P \approx \exp (\pi(r^2_1 + a^2)\Xi^{-1} + \pi(r^2_2 +
a^2)\Xi^{-1}).
\end{equation}
This is the exponential of  one quarter of the sum
of the outer and inner black hole horizon areas.

These two Fourier transformations are critical. Without them one
cannot even obtain the constrained gravitational instanton. 
The inclusion of the extra term due to the Fourier transformation
for the electrically charged rotating black hole pair also 
recovers the duality between the magnetic and electric cases
[3][8][9][10].

The construction of the constrained instanton using the inner and
outer black hole horizons is quite counter-intuitive. One
could also consider those constructions involving other horizons
as the instantons. However, the real part of the action for our
choice is always greater than that of the other choices for the
given configuration, and the wave function or the
probability is determined by the classical orbit
with the greatest real part of the action [1].

By the same argument, one has to use the pair of complex horizons
$r_0, r_1$ to construct the constrained instanton for the case of
open creation of black hole pair in the anti-de Sitter
background.  The relative probability of the Kerr-Newman black
hole pair creation takes a form similar to (13) with a
replacement of $r_1, r_2$ by $r_0, r_1$. One can show that the
sum of the four horizon areas is $24\pi/\Lambda$.
Therefore, one can
rewrite the relative probability as [8]
\begin{equation}
P \approx \exp -(\pi(r^2_2 + a^2)\Xi^{-1} + \pi(r^2_3 +
a^2)\Xi^{-1}).
\end{equation}
This is the exponential of the negative of one quarter of the sum
of the outer and inner black hole horizon areas.

It is interesting to note that the difference of relative
probabilities in the closed and open creations of black hole
pairs is the negative sign in the exponent. This is very
reasonable from a physical argument. Since for both cases, the
probability is a
decreasing function of the mass parameter.  This conclusion 
should be welcomed by quantum cosmologists.
 
The case of the Kerr-Newman black hole family with spatially
asymptotically flat infinity can be thought of
as the limit of our case as we let $\Lambda$ approach  $0$ from
below [8].

If one lets the angular momentum be zero, then it is
reduced into the Reissner-Nordstr$\rm\ddot{o}$m-(anti-)de Sitter
black hole case. If one further lets the charge be zero, then it
is  reduced into the Schwarzschild-(anti-)de Sitter black hole
case. There are only three horizons for the chargeless and
nonrotating case.

For the Schwarzschild-de Sitter black hole case, one has to use
the black hole and cosmological horizons to construct the
instanton, the creation probability is the exponential of the
entropy of the universe, or the exponential of one quarter of the
sum of the black hole and  cosmological horizon areas [3][11].
For the Schwarzschild-anti-de Sitter black hole case, one uses
the pair of complex horizons to construct the instanton, and the
creation probability is the exponential of the negative of
the  entropy. It is known that the entropy of the
Schwarzschild-anti-de Sitter universe is one quarter of the black
hole horizon area [12]. It is noted that the entropy is a
decreasing(increasing) function of the mass parameter for the
closed (open) model.

From  the no-hair theorem, all stationary black holes in the
de Sitter, anti-de Sitter and Minkowski spacetime backgrounds are
described by these Kerr-Newman families, so the problem of black
hole creations in these backgrounds is completely resolved. All
known  cases in the closed model are with regular
instantons [9][10][11][13][14][15], and they can be considered as
special cases of our study. The well known $S^4$ de Sitter
model without black hole and $S^2 \times S^2$ Nariai model with a
pair of maximal black holes have the maximal and minimal
creation probabilities, respectively [13].

Our treatment of quantum creation of the Kerr-Newman-anti-de
Sitter space family using the constrained instanton can
be thought of as a prototype of quantum gravity for an open
system, without appealing to the background subtraction approach
[16]. The beautiful aspect of our approach is that even in the
absence of a general no-boundary proposal for open universes, we
treat the creations of the closed and the open universes in the
same way.

It can be shown that  the probability of  the universe creation
without a black hole is greater than that with a pair of black
holes in all these backgrounds.

\vspace*{0.2in}  
 
\bf References:

\vspace*{0.1in}
\rm

1. J.B. Hartle and S.W. Hawking, \it Phys. Rev. \rm \bf D\rm
\underline{28}, 2960 (1983).

2. S.W. Hawking and N. Turok, \it Phys. Lett. \rm \bf B\rm
\underline{425}, 25 (1998), hep-th/9802030.

3. Z.C. Wu, \it Int. J. Mod. Phys. \rm \bf D\rm\underline{6}, 199
(1997), gr-qc/9801020.

4. Z.C. Wu, \it Gene. Relativ. Grav. \rm \underline{30}, 1639
(1998), hep-th/9803121.

5. N. Turok and S.W. Hawking, \it Phys. Lett. \rm \bf B\rm
\underline{432}, 271 (1998), hep-th/9803156.

6. Z.C. Wu,  \it Phys. Rev. \rm \bf D\rm
\underline{31}, 3079 (1985).

7. G.W. Gibbons and S.W. Hawking, \it Phys. Rev. \bf D\rm
\underline{15}, 2738 (1977).

8. Z.C. Wu, \it Phys. Lett. \bf B\rm  \underline{445}, 274
(1999); gr-qc/9810077.

9. S.W. Hawking and S.F. Ross,  \it Phys. Rev. \bf D\rm 
\underline{52}, 5865 (1995), hep-th/9504019.

10. R.B. Mann and S.F. Ross, \it Phys. Rev. \bf D\rm
\underline{52}, 2254 (1995), gr-qc/9504015.

11. R. Bousso and S.W. Hawking, hep-th/9807148.

12. S.W. Hawking and D.N. Page, \it Commun. Math. Phys. \rm
\underline{87}, 577 (1983).

13. R. Bousso and S.W. Hawking,  \it Phys. Rev. \bf D\rm
\underline{52}, 5659 (1995), gr-qc/9506047.

14. F. Mellor and I. Moss, \it Phys. Lett. \bf B\rm 
\underline{222}, 361 (1989).

15. I.J. Romans, \it Nucl. Phys. \bf B\rm 
\underline{383}, 395 (1992).

16. S.W. Hawking,  in \it General Relativity: An Einstein
Centenary Survey, \rm eds. S.W. Hawking and W. Israel, (Cambridge
University Press, 1979).
 
\end{document}